\documentclass[journal]{article}

\usepackage{arxiv}

\usepackage{hyperref}
\usepackage{graphicx}         

\usepackage{amsmath} 

\usepackage{amssymb}

\usepackage{amsthm}
\usepackage{color,soul}
\usepackage{flushend}
\usepackage{cite}
\usepackage{mathtools}
\usepackage{xcolor}
\usepackage{array,multirow}
\usepackage[ruled,linesnumbered]{algorithm2e}
\usepackage{wrapfig}

\allowdisplaybreaks
\newtheorem{remm}{Remark}
\newtheorem{assm}{Assumption}

\usepackage{tikz}
\usepackage{textcomp}
\usepackage{hyperref}
\usepackage{lipsum}

\newcommand\copyrighttext{%
  \footnotesize \textcopyright 2024 IEEE. Personal use of this material is permitted.
  Permission from IEEE must be obtained for all other uses, in any current or future 
  media, including reprinting/republishing this material for advertising or promotional 
  purposes, creating new collective works, for resale or redistribution to servers or 
  lists, or reuse of any copyrighted component of this work in other works. 
  DOI: \href{<http://tex.stackexchange.com>}{To be generated.}}
\newcommand\copyrightnotice{%
\begin{tikzpicture}[remember picture,overlay]
\node[anchor=south,yshift=10pt] at (current page.south) {\fbox{\parbox{\dimexpr\textwidth-\fboxsep-\fboxrule\relax}{\copyrighttext}}};
\end{tikzpicture}%
}

\usepackage{authblk}
\title{\Large \bf
Koopman Operator-based Detection-Isolation of \\ Cyberattack: A Case Study on Electric Vehicle Charging
}

\author[1]{Sanchita Ghosh}

\author[1]{Tanushree Roy}
\affil[1]{Department of  Mechanical Engineering, Texas Tech University, Lubbock, TX 79409, US. Emails:~{\tt\small sancghos@ttu.edu, tanushree.roy@ttu.edu}.}

\begin{document}

\copyrightnotice

\maketitle
\thispagestyle{empty}
\pagestyle{empty}

\begin{abstract}
One of the key challenges towards the  reliable operation of cyber-physical systems (CPS) is the threat of cyberattacks on system actuation signals and measurements. In recent years, system theoretic research has focused on effectively detecting and isolating  these cyberattacks to ensure proper restorative measures. Although both model-based and model-free approaches have been used in this context, the latter are increasingly becoming more popular as complexities and  model uncertainties in CPS increases. Thus, in this paper we propose a Koopman operator-based model-free cyberattack detection-isolation scheme for CPS. The algorithm uses limited system measurements for its training and generates real-time detection-isolation flags. Furthermore, we present a simulation case study to detect and isolate actuation and sensor attacks in  a Lithium-ion battery system of a plug-in electric vehicle during charging.

\end{abstract}

\section{INTRODUCTION}
Cyber-physical systems (CPSs) have become an integral part of the modern infrastructure attributing to their significant addition in  improved control and computational efficiency \cite{taheri2022data}. 
However, the operational reliability of CPSs hinges on the accurate sensor measurements, and  reliable actuation  which may be corrupted by cyberattacks \cite{pasqualetti2013attack}. Thus, much research has been focused on ensuring attack detection under both sensor  and actuation attacks \cite{roy2023actuator,lu2018secure,Troy_actuator_anomaly,ghosh2023security}. 
Authors in \cite{teixeira2012revealing,chen2017optimal,zhang2021stealthy} have proposed control-theoretic frameworks for various cyberattack policies and analyzed fundamental limitations in detecting them. 
In addition to detection, isolating the source of the attack is equally important to ensure mitigation and quick recovery after attack \cite{kanellopoulos2019moving}. In the context of cyberattack isolation, multiple observers-based filters have been designed to isolate actuation attacks and sensor attacks in \cite{zhang2022attack,ghosh2023cyberattack}. Similarly, a model-based data-driven isolation approach has been exploited in \cite{guo2017exploiting} for mobile robots. In these works \cite{zhang2022attack,ghosh2023cyberattack,guo2017exploiting}, complete model knowledge and considerable training data have been used for isolation.   However, reliable system model knowledge and sufficient training data are not always available for nonlinear systems with complex dynamics. Hence,  data-driven model-free approaches are often preferable for such systems \cite{bakhtiaridoust2022model}.

Model-free data-driven approaches based on the Koopman Operator (KO) have thus received a considerable amount of attention in recent years \cite{brunton2019notes,surana2016linear}. 
Unlike other model-free data-driven approaches such as machine learning, the Koopman analysis can identify the presence of anomaly in the case of unforeseen scenarios and insufficient training data \cite{bakhtiaridoust2022model,nandanoori2020model}. 
In \cite{gholami2022denoising} and\cite{nandanoori2020model}, authors focused on clustering of the Koopman Modes to identify the presence of anomalies and cyberattacks in system measurement, respectively. Moreover, \cite{nandanoori2020model} utilized embedding of spatial information in the Koopman modes to distinguish between sensor attacks and natural changes in system behavior. However, these model-free methods have not addressed the issue of isolation between the occurrence of an actuation  attack vs a sensor attack on the system. 
\newpage
To address these research gaps, our contributions in this work are as follows.
\begin{enumerate}
    \item We propose a KO-based real-time detection-isolation scheme for actuation and sensor attacks in CPS.
    \item The proposed scheme assumes no knowledge of the system model, requires only system measurements, and is trained online with limited data.
    \item We also derive the analytical conditions where actuation and sensor cyberattacks will be indistinguishable and thereby formalizing the fundamental limitations of the proposed isolation scheme.
    \item  Finally, we present a case study on compromised charging of a Lithium-ion battery system for a plug-in electric vehicle to illustrate the efficacy of our scheme.
\end{enumerate}

The rest of the paper is organized as follows: Section~\ref{preli} presents the theoretical background for Koopman analysis. Section~\ref{prob} explains the framework for this work and  Section~\ref{di_scheme} introduces the proposed detection- isolation scheme. Next, we present our simulation case study on the plug-in electric vehicles under cyberattack during charging in Section~\ref{simu}. Finally, Section~\ref{conclu} concludes our work.

\textbf{Notations:}
In this paper, 
$\dot{a}$ denotes the time derivative of the vector $a$; $|| \cdot ||_F$  denotes the Frobenious norm;  $A^\dag$ denotes the Moore-Penrose (Pseudo) inverse of the matrix $A$.

\section{PRELIMINARIES ON KOOPMAN OPERATOR} \label{preli}
In this section, we first present a brief overview of the KO analysis and then discuss a data-driven approach for the practical implementation of the operator. 
\subsection{Definition}
Let us first consider a CPS  such that the physical part exhibits  nonlinear dynamics defined as
\begin{align}\label{fx}
    \dot{x} = f(x(t),u(t));\quad y(t) = h(x(t)),
\end{align}
where $x  \in \mathbf{X} \subset \mathbb{R}^d$ is the state vector, $u  \in \mathbf{U} \subset \mathbb{R}^p$ is the control input,  $y \in \mathbb{R}^q$ is the output, and $h : \mathbf{X} \rightarrow \mathbb{R}^q$ denotes the nonlinear output function. Here the system dynamics $f: \mathbf{X} \times \mathbf{U} \rightarrow \mathbb{R}^d$ is a (continuously differentiable) nonlinear function and the flow map $\Phi^t(x,u)$ is a solution of the ODE \eqref{fx} 
for the time between $[0,t]$. 
Our primary objective in using KO theory is to obtain an \textit{ linear, infinite-dimensional} state dynamics that approximate the \textit{nonlinear, finite-dimensional} state dynamics in \eqref{fx} \cite{koopman1931hamiltonian}. Additionally, data-driven approaches towards approximating the KO have enabled us to learn the dynamics of systems with uncertainty, nonlinearity, and complexity using computationally efficient algorithms and limited data (vide references in \cite{brunton2019notes}).

To define the KO, let us  consider an infinite-dimensional Hilbert space of observable functions $\mathcal{F}$. Furthermore, since the boundedness of the KO is not guaranteed, we assume that our KO has infinitely many discreet eigenvalues corresponding to infinitely many eigen-observables.  Then the set of KO on this space of observables, $\mathcal{K}^t : \mathcal{F} \rightarrow \mathcal{F}$ is defined using the Koopman eigenfunctions (KEF), $\phi \in \mathcal{F} $, and the corresponding Koopman eigenvalues (KEV), $\lambda \in \mathbb{C}$, as \cite{bruder2021advantages}:
\begin{align} \label{kt_phi}
    \left[ \mathcal{K}^t \phi \right] (x,u) = \phi\left(\Phi^t(x,u) \right)= e^{\lambda t} \phi (x,u); 
\end{align}
The primary significance of \eqref{kt_phi} lies in the fact that the KO evolves the KEFs linearly. 
Now, we can consider ${\hat{\psi}}(x) \in \mathbb{C}^k$ be any vector-valued observable function that lies on the space  $\mathcal{F} =\text{span} \{\phi_i \}_{i = 1}^\infty$ where the KEFs $\phi_i: \mathbf{X} \rightarrow \mathbb{C}$ are the eigen-observables.  Then  ${\hat{\psi}}(x)$  can be expanded in terms of KEFs as $
    {\hat{\psi}}(x) = \sum_{i=1}^{\infty} \phi_i v_i^{\hat{\psi}}.$
Here the Koopman Modes (KM) $v_i^{\hat{\psi}} \in \mathbb{C}^k$ are the coefficients of the projection of ${\hat{\psi}}(x)$ onto the span$\{\phi_i\}_{i = 1}^\infty$  \cite{mezic2005spectral}.  Furthermore, such expansion in terms of KEFs and KMs is referred to as Koopman Mode Decomposition (KMD). 

\begin{remm} \label{vtor}
    It should be noted from  \eqref{kt_phi}, that the KEFs and KEVs are solely dependent on the system dynamics and the function space $\mathcal{F}$. On the other hand, the KMs $v_i^{\hat{\psi}}$ are specific to the observable ${\hat{\psi}}(x)$. 
\end{remm}

 Next, using assumptions in \cite{brunton2019notes}, if the observable function ${\hat{\psi}}(x)$ lies in the subspace spanned by the finite set of KEFs, a good approximation can be achieved by ${\hat{\psi}}(x) = \sum \limits_{i=1}^{n} \phi_i v_i^{\hat{\psi}}$. 
Moreover, we can assume that the measurements of the system \eqref{fx} are special observable functions \cite{surana2016linear}. This is mathematically expressed in the following assumption.
\begin{assm}\label{kmd}
    There exist a finite subset of KEFs $\phi_i (x), \forall i \in \{1, 2, \cdots, n\}$ for $n> d$, such that the system \eqref{fx} can be represented as $ y = h(x) = \sum \limits_{i=1}^n \phi_i (x) v_i^h.$
    In other words, we can estimate a finite subset of KEFs such that the $h(x)$ lies in the subspace $\mathcal{F}^n = \text{span} \{ \phi \}_{i=1}^n$. Here $v_i^h \in \mathbb{C}^q$ are the KMs asSOCiated with the measurements.
\end{assm}
\subsection{Implementation}
Delay Embedding is a data-driven approach where Taken's theorem is exploited to obtain a reliable approximation of system states from a consecutive measurement data sequence \cite{kamb2020time}. This  also provides a means to learn a lower dimensional Koopman invariant space and the corresponding KMD \cite{arbabi2017ergodic}. We consider here that only the measurement and input data matrices are available and
 we arrange them  as follows. \vspace{-2mm}
\begin{align} \label{hdmds}
    &Y_b = \begin{bmatrix}
        D_1 & D_2 & \cdots & D_{m-\tau-1}
    \end{bmatrix},\\ &Y_s = \begin{bmatrix}
        D_2 & D_3 & \cdots & D_{m-\tau}
    \end{bmatrix}, \\ & U_b = \begin{bmatrix}
        u_{1+\tau} & u_{2+\tau} & \cdots & u_{m -1}
    \end{bmatrix}.
\end{align} Here $D_l = \begin{bmatrix}
    y_l^T & u_l^T & y_{l+1}^T & u_{l+1}^T & \cdots & y_{l+\tau}^T
\end{bmatrix}^T$; $y_l \in \mathbb{R}^q$ and $u_l \in \mathbb{R}^p$  for the system \eqref{fx} at the $l$-th time instant; $\tau$ is the embedded delay. Next, we assume that for sufficiently embedded data matrices, we can obtain Koopman linear approximation for the system \eqref{fx} from the optimization problem posed below.
\begin{align}
    Y_s \approx A Y_b + B U_b,& \label{lin_y} \\
    \min\limits_{\Lambda} || Y_{s} -  \, \Lambda \Upsilon||_F, \,\,\,
    \Upsilon = \begin{bmatrix}
    Y_{b} &
    U_b
\end{bmatrix}^T, &\,\,\, \Lambda = \begin{bmatrix}
    A & B
\end{bmatrix}.\label{opt_y}
\end{align}
The analytical least-square solution of \eqref{opt_y}
can be found as $\Lambda = Y_{s} \Upsilon^\dag$.
 Notably, for our proposed scheme, the delay method is applied over a sliding window to first learn the model and subsequently predict it over a receding horizon. Hence, we define a moving window sequence of $W$ observations and for $  \Tilde{W} < W$ we split the window into two sub-sequence: 
$\text{Learning window:} \, \, \quad \mathfrak{L} \in \{ s-W, \cdots, s-\Tilde{W}\}, $ and $\text{Prediction window:} \quad \mathfrak{P} \in \{ s- \Tilde{W}+1, \cdots, s\}$.
 We note here that the learning window $\mathfrak{L}$ is larger than the prediction window $\mathfrak{P}$ and the sliding window is moved ahead with $\Tilde{W}-1$ amount after every prediction window. Moreover, $W-\Tilde{W}$ is the $m$ from \eqref{hdmds}. With these preliminaries on KO theory and the methods of its numerical computation, we are now ready to discuss the framework of our problem.

\section{PROBLEM FRAMEWORK} \label{prob}
Let us consider the nonlinear system \eqref{fx} under cyberattack such that this cyberattack can be either an actuation attack $\delta_u$ or a sensor attack $\delta_y$. The attacked physical part of the CPS is then given by:
\begin{align}\label{fx_attack}
    \dot{x} = f(x(t),u(t)+\delta_u);\quad y(t) = h(x(t))+\delta_y.
\end{align}
Here we assume that the system model is unknown and only the system measurements $y$ and the control input signal $u$ are available to us. Our objective is to detect and isolate the presence of these attacks $\delta_u$ and $\delta_y$.

\begin{figure}[t!]
    \centering
    \includegraphics[trim = 0mm 0mm 0mm 0mm, clip,  width=0.75\linewidth]{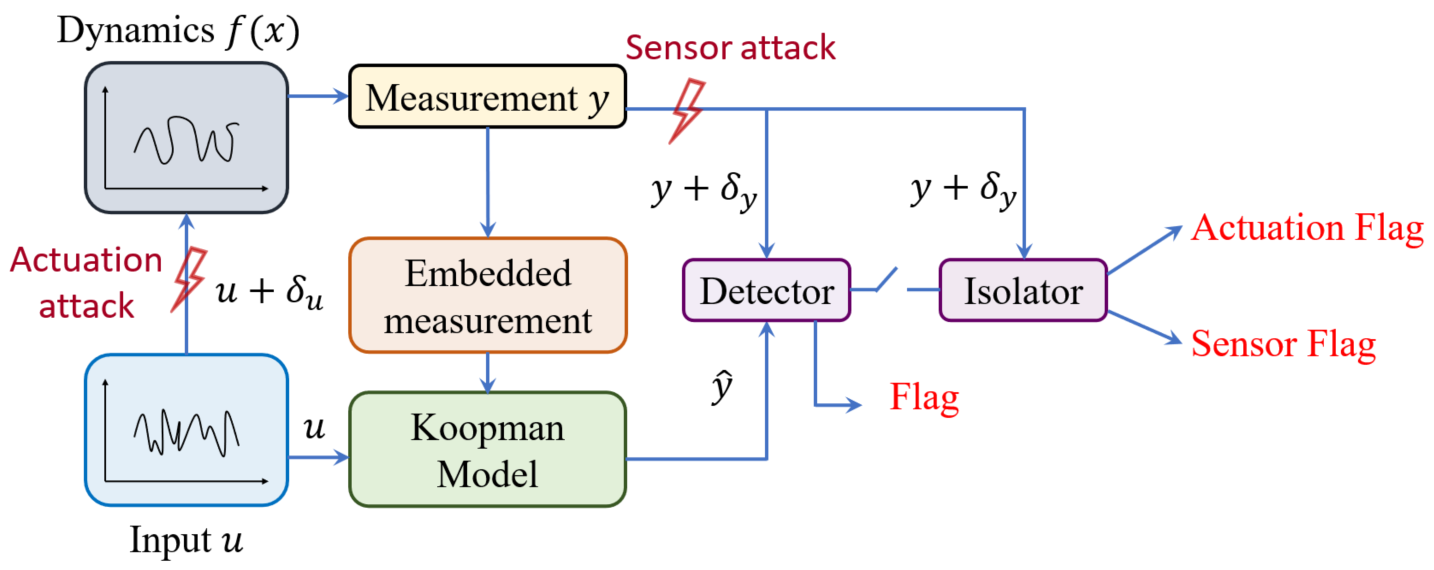}   
    \caption{A block diagram showing the Koopman approximation-based detection-isolation scheme for a nonlinear CPS.}
    \label{fig:model}
\end{figure}

\begin{assm}\label{sep_att}
    In this framework, stealthy attacks such as zero dynamic attacks or covert attacks which simultaneously manipulate actuation and measurements are not considered because of their inherent undetectability using residual-based detection systems \cite{pasqualetti2013attack,roy2020secure}. Similarly, sensor attacks that can spoof the behavior of actuation attacks are also not considered for the isolation scheme.
\end{assm}
To achieve our objective, we utilize \eqref{lin_y}-\eqref{opt_y} to obtain the measurement prediction $\hat{y}$. Furthermore, we assume there is a prediction window $\mathfrak{P}$ over which the predictions are based on the uncompromised system learning while the corresponding measurements are from the compromised system. Therefore, a residual-based detection filter can capture the presence of cyberattacks in the prediction window $\mathfrak{P}$. In particular,

following Assumption~\ref{kmd}, we have \vspace{-2mm}
\begin{align}\label{y} 
\hat{y} = \sum \limits_{i=1}^n \phi_i (x) v_i^h.  
\end{align} Then,  $\hat{y}$ is sent to the detection-isolation scheme to detect and isolate the presence of cyberattacks $\delta_u$ and $\delta_y$ as shown in the block diagram of the proposed scheme for a nonlinear CPS in Fig~\ref{fig:model}. From \eqref{y}, it is evident that if either KEFs $\phi_i$ or the KMs $v_i^h$ of the system are changed, then the predicted $\hat{y}$ will be different from the measured $y$. At the same time, it is clear from \eqref{fx_attack} that the actuation attack explicitly affects the system dynamics while sensor attack affects the system measurements or the observable functions. Consequently, referring to Remark~\ref{vtor}, we can posit that actuation attack 
affects the KEFs $\phi_i$ part and the sensor attack affects the KMs $v_i^h$ part of the system measurement. With this notion, we will present our proposed scheme.

\section{DETECTION- ISOLATION SCHEME} \label{di_scheme}
The KO-based detection-isolation scheme consists of three main steps. Generation of (i)  an approximate linear model of the system, (ii)  an \textit{attack detection} decision a flag, (iii)  an appropriate actuation or sensor \textit{attack isolation} flag. 
\subsection{Detection scheme}
Now, we can assert that there exists a prediction window $\mathfrak{P}$ during which the system \eqref{fx} is under an actuation attack or a sensor attack. Consequently, during this prediction window $\mathfrak{P}$, we can represent the corrupted system measurement using either the modified KEFs $\Tilde{\phi}_i (x)$ for actuation attacks or the modified KMs $\Tilde{v}_i^h$ for sensor attacks.  The corrupted measurement is then given by:
\begin{equation} \label{att_scn}
  {y} = \begin{cases}
   \sum \limits_{i=1}^n \Tilde{\phi}_i (x) v_i^h, & \text{if actuation attack}; \\
   \sum \limits_{i=1}^n \phi_i (x) \Tilde{v}_i^h, & \text{if sensor attack}. 
\end{cases}
\end{equation}

Meanwhile,  no attack is present in that learning window, and thus our prediction will solely be based on our learning of our nominal system model. This prediction is given by \eqref{y}.
Now, using the actual and predicted measurement \eqref{att_scn}-\eqref{y}, we define our detection residual as $ r_D = y-\hat{y}$. For asyncrnous actuation and sensor attacks satisfying Assumption~\ref{sep_att}, this residual is given by
\begin{equation}\label{res_df}
   r_D = \begin{cases}
\sum \limits_{i=1}^n \left(\Tilde{\phi}_i  - {\phi_i} \right) v_i^h,  & \text{if actuation attack}; \\
 \sum \limits_{i=1}^n   \phi_i \left( \Tilde{v}_i^h - {v}_i^h \right), & \text{if sensor attack}.
\end{cases}
\end{equation}
We note here, we droped $x$ from the variable $ \phi_i (x)$ for the brevity of notation. Now, it is evident from \eqref{res_df} that in the absence of any attack, the KEFs and KMs remain unchanged i.e. $\Tilde{\phi}_i - {\phi}_i =0$ and $\Tilde{v}_i^h - {v}_i^h=0$. This implies that the residual $r_D$ \eqref{res_df} will remain close to zero under no attack.

Our detection scheme utilizes this to make a no-attack decision, and when the residual $\|r_D\|$ first crosses a predefined threshold, the attack flag is turned on. This threshold is set above the fluctuations of the residual signal under nominal system conditions \cite{Ding}. Additionally, we note here that this residual does not continue to cross the threshold after the next learning window as the algorithm starts to assimilate the attack signature into the system model.  However, the residual again crosses the threshold once the attack has been removed from the system. Thus,  the threshold crossing is utilized to generate (turn on or off) an attack flag.

\subsection{Isolation scheme}
Upon detection of a cyberattack, the proposed isolator is  activated to ascertain if the attack is injected via actuation  or measurement.

 This isolation scheme is based on the behavior of the residual signal \eqref{res_df}. In particular, we know that during actuation attack the generated detection residual $r_D=\sum \limits_{i=1}^n \Delta \phi_i  v_i^h$, where  $\Delta \phi_i = \left[ \Tilde{\phi}_i   -  {\phi}_i  \right]$. Alternatively, during sensor attack, the detection residual $ r_D = \sum \limits_{i=1}^n \phi_i  \Delta v_i$, where $\Delta v_i = \left[ \Tilde{v}_i^h -  {v}_i^h \right]$. This implies that $r_D$ is either a linear combination of $ \{\Delta{\phi}_i\}$ with weights $v_i^h$ during an actuation attack or a linear combination of $\{{\phi}_i\}$ with weights $\Delta v_i^h$ during sensor attack.

 However, we note here that the adversary can craft a sensor attack such that the output of the system is indistinguishable from a case of an actuation attack. Such actuation-attack-spoofing sensor attacks will satisfy:\vspace{-2mm}
\begin{align}\label{spoof}
     y = \sum \limits_{i=1}^n \Tilde{\phi}_i (x) v_i^h =
   \sum \limits_{i=1}^n \phi_i (x) \Tilde{v}_i^h.
\end{align}
Thus, their signature on the detector residual will be identically equal to $
     r_D = \sum \limits_{i=1}^n \phi_i  \Delta v_i = \sum \limits_{i=1}^n \Delta \phi_i  v_i^h.$ Hence, these actuation-attack-spoofing sensor attacks will remain undetected and \eqref{spoof} gives the analytical condition on the measurement which creates 
 \textit{fundamental limitation on this measurement-based isolation scheme.} By Assumption~\ref{sep_att}, we do not consider such non-isolable attacks in this work.

 Consequently, to isolate between actuation and sensor attacks satisfying Assumption~\ref{sep_att}, we check whether we can find a set of vectors ${\widehat{\Delta\phi}_i }$ whose linear combinations using the unchanged KMs $v_i^h$ match with the residual $r_D$. If such basis vectors are found, then the $r_D$ is generated due to actuation attacks (since KMs are unchanged) and it is generated by sensor attack otherwise.
 Exploiting this idea, we define our isolation residual $r_I(t)$ as the following optimization: \vspace{-2mm}
\begin{align}\label{opt}
   r_I =  \min_{\widehat{\Delta\phi}_i } \left\| \sum\limits_{i=1}^n \widehat{\Delta\phi}_i v_i^h - r_D \right\|. 
\end{align} 
Essentially, this optimization \eqref{opt} tries to find the basis for the residual signal $r_D$ with the unchanged weights or KMs $v_i^h$ over the prediction window. This implies that for some small positive $\epsilon \in \mathbb{R}$, we can define our isolation rule as: $ 0 \leqslant r_{I} \leqslant \epsilon \quad\text{implies actuation attack}, $ and $ r_{I} > \epsilon >0 \quad \text{implies sensor attack.}$

Next, we prove the above intuition mathematically for two cases of actuation and sensor attack.

\noindent
\textbf{Case 1   (Actuation attack):}
For this case, $r_D=\sum \limits_{i=1}^n \Delta \phi_i  v_i^h$ and replacing this $r_D$ in the definition of $r_I$ \eqref{opt} we get

\begin{align}\label{r_I_act}
    r_I &=   \min_{\widehat{\Delta\phi}_i } \left\| \sum\limits_{i=1}^n \widehat{\Delta\phi}_i v_i^h - \sum\limits_{i=1}^n \Delta \phi_i  v_i^h \right\| .
\end{align}
Furthermore, rearranging the terms and noting that the minimization can be obtained for $\widehat{\Delta\phi}_i = \Delta \phi_i$ we can deduce
\begin{align}
    r_I  =  \min_{\widehat{\Delta\phi}_i } \left\| \sum\limits_{i=1}^n \left[ \widehat{\Delta\phi}_i - \Delta \phi_i \right] v_i^h \right\| \leqslant \epsilon. \label{act_ep}
\end{align}

\noindent
\textbf{Case 2   (Sensor attack):}
Conversely, in the case of a sensor attack: \mbox{ $ r_D = \sum \limits_{i=1}^n \phi_i (x) \Delta v_i $}. Now, inserting this expression into the definition of $r_I$ \eqref{opt} and rearranging the terms:
\begin{align}
     r_I &=   \min_{\widehat{\Delta\phi}_i } \left\| \sum\limits_{i=1}^n \left[ \widehat{\Delta\phi}_i + \phi_i  \right]  {v}_i^h - \sum \limits_{i=1}^n \phi_i    \Tilde{v}_i^h  \right\|. \label{ses_ri}
\end{align} 
Using reverse triangle inequality for \eqref{ses_ri} yields:
\begin{align} \label{ses_ri2}
    r_I &\geqslant  \min_{\widehat{\Delta\phi}_i } \left|\, \left\|\sum\limits_{i=1}^n \left[ \widehat{\Delta\phi}_i + \phi_i  \right]  {v}_i^h \right\| - \left\|\sum\limits_{i=1}^n  \phi_i   \Tilde{v}_i^h \right\|\,\right| .
\end{align}
We note here that the second term in \eqref{ses_ri2} is $y(t)$. Since Assumption~\ref{sep_att} is true, from \eqref{spoof} we claim that there exists no $\Tilde{\phi}_i (x)$ such that $\sum \limits_{i=1}^n \Tilde{\phi}_i (x) v_i^h =   \sum \limits_{i=1}^n \phi_i (x) \Tilde{v}_i^h$. This in turn implies that the optimization in \eqref{ses_ri2} is not able to find a $\widehat{\Delta\phi}_i$ such that  $\left\| \sum\limits_{i=1}^n \left[ \widehat{\Delta\phi}_i + \phi_i (x) \right]  {v}_i^h \right\| = \left\|  \sum \limits_{i=1}^n \phi_i (x)   \Tilde{v}_i^h  \right\|$. Moreover,  by Assumption~\ref{sep_att}, $y(t) \not\equiv 0$ as this will imply a zero-dynamics attack. Thus, these two restrictions on \eqref{ses_ri2} guarantee that $r_I\neq 0$. Consequently, this proves that for some $\epsilon>0,$ the residual $r_I>\epsilon$ for the sensor attack case.
\vspace{2mm}

\textbf{Choice of threshold for isolation residual:} The choice of threshold for the isolation scheme relies on the choice of $\epsilon$. This number can be determined by finding the isolation residual of the system under maximum actuation attack capacity i.e. $\|\delta_u\|_{max}\leqslant \|u_{max}\|.$

\section{A CASE STUDY ON  ELECTRIC VEHICLE  CHARGING }\label{simu}

In this section, we present the simulation case study on cyberattack scenarios on a plug-in electric-vehicle (PEV) through the electric vehicle supply equipment (EVSE) at charging stations. A recent report from Sandia National Lab on the risk assessment of EVSE created a threat model and clearly indicated how the attacks on automotive charging will be exacerbated as more PEVs are equipped with extreme fast charging, DC fast charging, etc \cite{johnson2022cybersecurity}. 

\begin{figure}[ht!]
    \centering
    \includegraphics[trim = 0mm 15mm 0mm 9mm, clip,  width=0.6\linewidth]{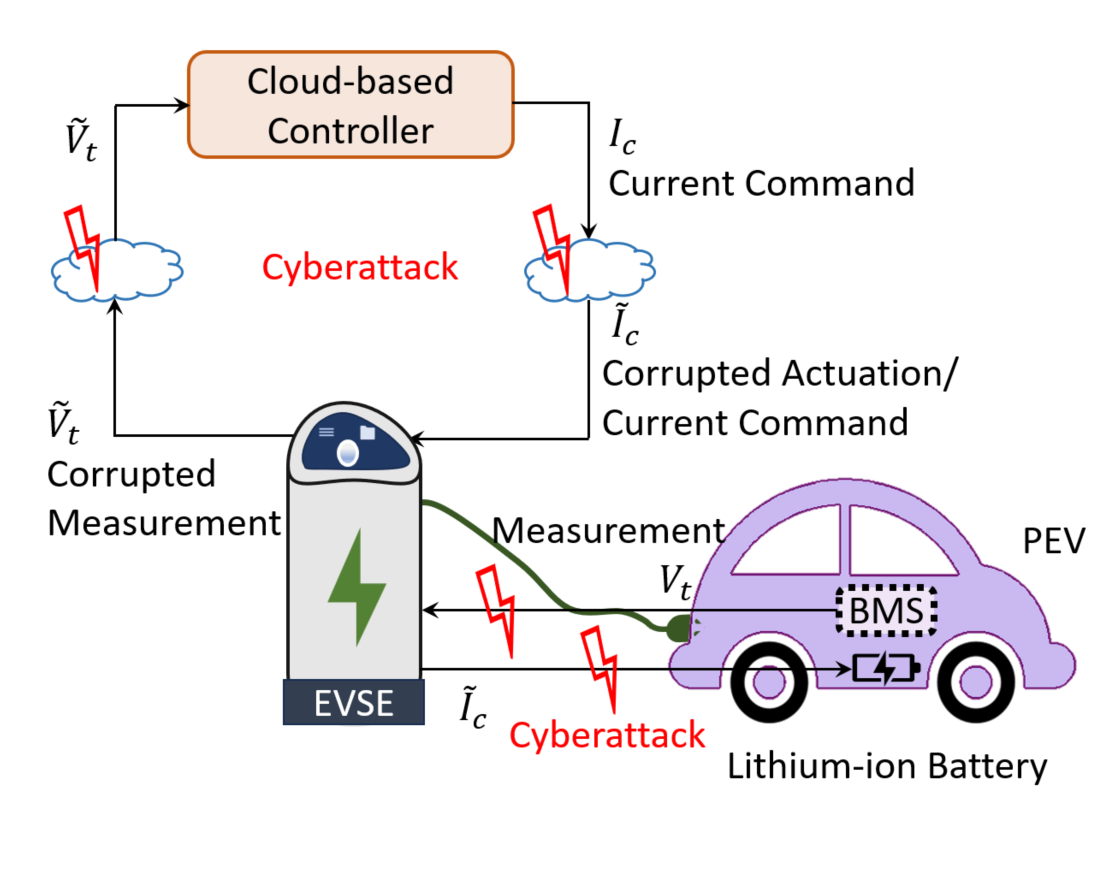}   
    \vspace{-2mm}
    \caption{ Potential cyberattack vectors between the cloud-based charging control, EVSE, and PEV.}
    \label{fig:model_batt}
\end{figure}

For this case study, following \cite{johnson2022cybersecurity} we will consider: (i) the corruption of communication between the electric vehicle communication controller (EVCC) and supply equipment communication controller (SECC), (ii) the corruption of the battery management system  (BMS) and charger firmware, and (iii) corruption of communication between EVSE and charging station cloud-based control.
 Here, the terminal voltage measurement of the PEV battery is sent from the BMS to the EVCC and finally to the cloud controller. Corruption of this measurement along this communication is considered as a sensor attack. Similarly, charging current policies are sent from the cloud controller to the EVCC and finally to BMS. The corruption of this current command is our actuation attack. Figure~\ref{fig:model_batt} shows these potential cyberattack vectors between the cloud-based charging control, EVSE, and PEV. 

For this study, we generated the terminal voltage measurements for the battery of the PEV using a nonlinear equivalent circuit model for a prismatic cell Li-ion battery \cite{samad2014parameterization}:
\begin{align}\label{X_batt}
   & \quad \quad \quad \quad \quad \quad \dot{X} = A(X) X + B(X) I_c, \quad \text{where} \\
    &  X = \begin{bmatrix}
        \dot{V_1} &
        \dot{V_2} &
        \dot{\xi} 
    \end{bmatrix}^T,  \quad B(X) =  \begin{bmatrix}
        \frac{1}{C_1(\xi)} &
        \frac{1}{C_2(\xi)} &
        -\frac{1}{Q}
    \end{bmatrix}^T, \nonumber \\ 
    & A(X) = \text{diag}\left(- [{R_1(\xi) C_1(\xi)}]^{-1}, - [{R_2(\xi) C_2(\xi)}]^{-1} ,0\right). \nonumber
\end{align} Here $\xi$ and $Q$ denote the state of charge (SOC) and the nominal capacity of the battery, respectively. The states $V_1$ and $V_2$ are the voltages across the two resistance capacitance pair, representing the diffusion of lithium in the solid and electrolyte. The SOC dependent resistances $R_1, R_2$ and capacitances $C_1, C_2$ in \eqref{X_batt}  are defined as $ R_i = R_{i0} + R_{i1} \xi + R_{i2} \xi^2,\, \text{and}\,\, C_i = C_{i0} + C_{i1} \xi + C_{i2} \xi^2 +T \left( C_{i3} + C_{i4} \xi +C_{i5} \xi^2 \right)$ for $i\in \{1,2\} $.

The parameters  are given in Table~\ref{tab:RC}. We also note here that $T=298^{\circ} K$ represents the ambient temperature.  Finally, the measured terminal voltage $V_t$ of the battery is defined in terms of the system states as $ V_t = OCV - I_c R_s - V_1 - V_2$,
where OCV denotes the open circuit voltage of the battery which has a strong dependence on the SOC instant of the battery and can be obtained from OCV-SOC plot   \cite{samad2014parameterization}. Furthermore, $I_c < 0 $ specifies charging, and $I_c > 0 $ specifies discharging scenario. The PEV is charged with a constant current constant voltage (CCCV) policy and the charging cycle is generated by the cloud-based controller of the charging station.  Additionally, the rated battery capacity is assumed to be $Q = 5 Ah $ and the internal series resistance is considered as \mbox{$R_s = 0.0048 \Omega$}. 

We note here that our analysis is based on the model of a battery cell, instead of a battery pack. However, our work can be extended to the battery pack by cascading multiple battery cells \cite{dey2020cybersecurity}. This will be addressed in our future work.

\begin{wraptable}{r}{7.5cm}
\setlength{\tabcolsep}{5pt} 
\renewcommand{\arraystretch}{1} 
\begin{tabular}{|*{2}{c|}}
  \hline
  $R_{10} = 0.0701135 \mu \Omega$, & $R_{20} = 0.0288 \Omega$,\\ \hline
     $R_{11} = - 0.043865 \mu \Omega$,&  $R_{21} = - 0.073 \Omega$,\\ \hline
     $R_{12} = 0.023788 \mu \Omega$,  &$R_{22} = 0.0605  \Omega$,\\ \hline
    $C_{10} = 335.4518 F$,  & $C_{20} = 31.881 kF$, \\ \hline
     $C_{12} = -1.3214 kF$, & $C_{22} = 104.93 kF$, \\ \hline
    $C_{14} = - 65.4786 F/^{\circ} K$, & $C_{24} = 10.1755 kF/^{\circ} K$,\\ \hline
     $C_{11} = 3.1712 kF,$  &$C_{21} = - 115.93 kF$, \\ \hline
       $C_{13} = 53.2138 F/^{\circ} K,$ &$C_{23} = 60.3114 F/^{\circ} K$, \\ \hline
      $C_{15} = 244.3761 F/^{\circ} K,$ &$C_{25} = -9.5924 F/^{\circ} K$. \\
  \hline
\end{tabular}
\vspace{1mm}
\caption{List of parameters required in \eqref{X_batt}.}
    \label{tab:RC}
\end{wraptable}

The PEV is considered to start charging with $35\%$ SOC with a $5 A$ current under the CCCV policy. Furthermore, we consider a control protocol to charge the PEV up-to $94\%$ SOC, since typically  PEVs are not charged fully \cite{dey2020cybersecurity}.
To implement our detection-isolation scheme for this PEV charging scenario, we adopted   the delay embedding based data-driven approach for Koopman-operator approximation. We considered a moving window of $W = 23$ snapshots in time with the embedded delay  $\tau = 13$ \eqref{hdmds}. This moving window consisted of a learning window of 15, followed by a prediction window of 8.
Then, using the nominal terminal voltage measurement data, we define our detection threshold as $0.002$ \cite{Ding}.  To define the isolation threshold, we plotted the isolation residual with maximum actuation attack capacity for this battery, which we consider to be  5C rate charging current or $25A$. Considering the maximum isolation residual for an actuation attack for  $25A$, we defined the isolation threshold as $0.12$. Next, we will present the two case studies of actuation attack and sensor attack scenarios on this battery and demonstrate the efficacy of our scheme. 

\subsection{Case I: Actuation attack}
\begin{figure}[h!]
    \centering
    \includegraphics[trim = 0mm 0mm 0mm 0mm, clip,  width=0.5\linewidth]{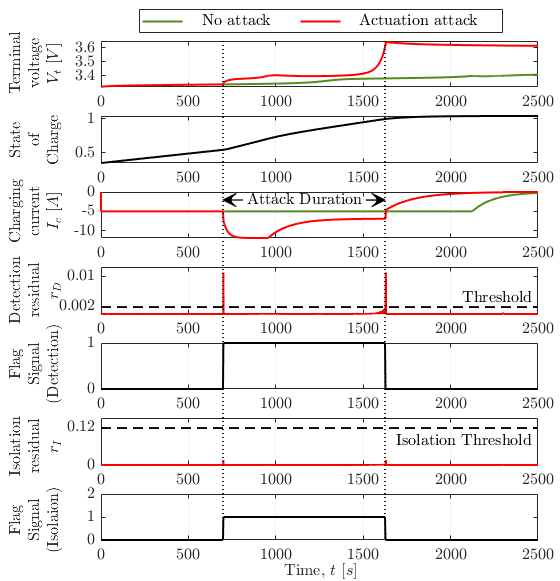}   
    \caption{Plot shows the nominal and corrupted terminal voltage, SOC, charging current, detector and isolator residuals for actuation attack.}
    \label{fig:det_att}
\end{figure}
In this scenario, we consider that the input signal $I_c$ is corrupted by the adversary to achieve overcharging. The battery model under  attack can be obtained from \eqref{X_batt} as
\begin{align}
    & \dot{\Tilde{X}} = A(\Tilde{X}) \Tilde{X} + B(\Tilde{X}) \left(I_c + \delta_u \right), \quad \Tilde{V}_t = OCV - \left(I_c + \delta_u \right) R_s - \Tilde{V}_1 - \Tilde{V}_2.
\end{align}
To achieve overcharging, a high charging current  is given to the battery, between  $700s$ and $1600s$ (shown in $3^{\text{rd}}$ plot of \mbox{Fig. \ref{fig:det_att}}). This causes the battery to overcharge, as is evident from $V_t$ (top plot) and SOC ($2^{\text{nd}}$ plot) of \mbox{Fig. \ref{fig:det_att}}. Now, from the $4^{\text{th}}$ plot in \mbox{Fig. \ref{fig:det_att}}, we observe that the detector residual  crosses the threshold twice -- once when the attack begins and once when the attack ends. Hence, the detection flag is set to 1 between the subsequent threshold crossings (shown in $5^{\text{th}}$ plot of \mbox{Fig. \ref{fig:det_att}}). The detection occurs within $1s$ of the attack, which shows the efficiency of our detection algorithm. Next, we analyze the performance of our isolation scheme. From the $6^{\text{th}}$ plot of \mbox{Fig. \ref{fig:det_att}}, we observe that the isolation residual remains below the isolation threshold for actuation attack, and the isolation flag in set to 1 in the last plot of \mbox{Fig. \ref{fig:det_att}}. Thus, the scheme correctly identifies the detected attack is an actuation attack, and achieves cyberattack isolation.

\subsection{Case II: Sensor attack}
In this scenario, the adversary alters the measured terminal voltage $V_t$,  without corrupting the charging current command $I_c$ generated by the cloud-based controller. Hence, the corresponding battery model can be defined as
\begin{align}
    & \dot{X} = A(X) X + B(X) I_c, \quad \Tilde{V}_t = OCV - I_c R_s - V_1 - V_2 + \delta_y. \label{xsens}
\end{align}\vspace{-5mm}
\begin{figure}[h!]
    \centering
    \includegraphics[trim = 0mm 0mm 0mm 0mm, clip,  width=0.5\linewidth]{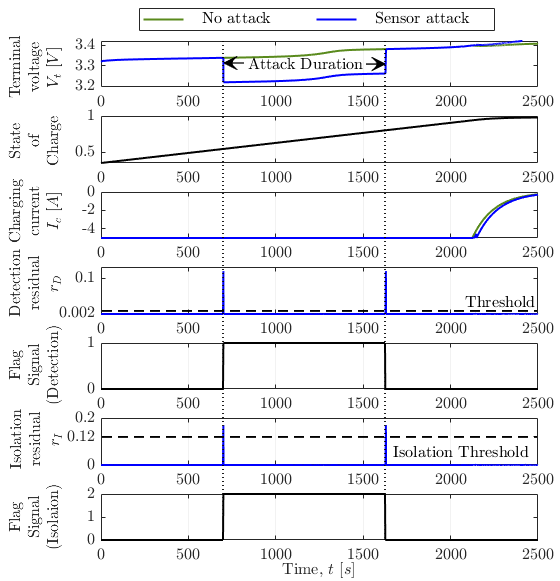}   
    \caption{Plot shows the nominal and corrupted terminal voltage, SOC, charging current, detector and isolator residuals for sensor attack.}
    \label{fig:det_sen}
\end{figure}

The adversary can force the battery to overcharge by corrupting the sensor data and communicating lower $V_t$ to the controller, since  the battery charging protocol is based on $V_t$. 
We simulated this particular scenario by injecting a false voltage data to the controller from $700s$ to $1600s$ (shown in top plot of \mbox{Fig. \ref{fig:det_sen}}).  Due to this terminal voltage corruption, the $2^{\text{nd}}$ plot of \mbox{Fig. \ref{fig:det_sen}} shows the SOC of the battery exhibiting eventual overcharging of the cell. Additionally, such sensor attack delays the constant voltage charging of the battery as shown in the $3^{\text{rd}}$ plot of \mbox{Fig. \ref{fig:det_sen}}. Under this sensor attack scenario,  the detection residual crosses the threshold ($4^{\text{th}}$ plot) in the beginning and the end of the attack injection. The detection flag is thus set to 1 between these two crossing to indicate the presence of a cyberattack. The generated detection flag is shown in the $5^{\text{th}}$ plot of \mbox{Fig. \ref{fig:det_sen}}. After detection, the isolation scheme is activated and the isolation residual $r_I$ is calculated. For this case, we observe that the isolation residual crosses the isolation threshold (as shown in the $6^{\text{th}}$ plot). Accordingly, the isolation flag is set to 2 to indicate the presence of sensor attack in the system (as shown in the $7^{\text{th}}$ plot of \mbox{Fig. \ref{fig:det_sen}}).  Thus, we validate the performance of the isolation scheme.

\newpage
\section{CONCLUSIONS}\label{conclu}

In this work, we have used a Koopman operator-based model-free approach for the detection and isolation of actuation and sensor attacks in CPS. We have utilized  delay embedding method to show that signature of the actuation and sensor attack on the Koopman operator modes and eigenfunctions can be used to detect and isolate cyberattacks in CPS. Most importantly, this detection-isolation scheme uses only system measurement and learns the system model with limited data. This is important for faster and efficient generation of attack flags in presence of cyberthreats. Additionally, we have derived conditions where such isolation scheme will fail owing to the presence of actuation-attack-spoofing sensor attack.  We also presented a case study of cyberattack on a Lithium-ion battery powered PEV, charged by EVSE. The simulation results show that our proposed scheme can detect and isolate the presence of corrupted charging policies (actuation attack) and corrupted measurement (sensor attack).




\bibstyle{arxiv}
\bibliography{ref.bib,Ref_Troy}

\end{document}